\documentclass [10pt,twoside]{article}

\usepackage{shrthnds}
\usepackage{cite}
\usepackage{graphicx}
\usepackage[usenames]{color}
\usepackage{subfigure}
\usepackage{setspace}

\usepackage{multirow}

\usepackage[hang,small]{caption}

\usepackage{amsmath}                  

\usepackage{geometry}
\usepackage{fancyhdr}
\usepackage{titlesec,titletoc}
  \titleformat{\section}{\Large\sf\bfseries}{\thesection}{1em}{}
  \titleformat{\subsection}{\large\sf\bfseries}{\thesubsection}{1em}{}

\title{\sf\bfseries \ntitle}
\author{\normalsize  Pankaj Jain$^{1}$\footnote{email: pkjain@iitk.ac.in}~,
Purnendu Karmakar$^1$\footnote{email: purnendu@iitk.ac.in}~,
 Subhadip Mitra$^{2}$\footnote{email: smitra@iopb.res.in}~,\\ Sukanta
Panda$^3$\footnote{email: sukanta@iiserbhopal.ac.in} and Naveen K. Singh$^1$\footnote{email: naveenks@iitk.ac.in}}
\date{}

\newcommand{\pghdr}{\footnotesize {P. Jain} {\it et al.} -- Cosmological Perturbation Analysis in a Scale Invariant Model of Gravity}
\newcommand{\ntitle}{Cosmological Perturbation Analysis in a Scale Invariant Model of Gravity}

\begin{document}
\vspace{-3cm}
\maketitle
\vspace{-0.6cm}
\bc
{\small 1) Department of Physics, IIT Kanpur, Kanpur 208 016, India\\
2) Institute of Physics, Bhubaneswar 751 005, India\\
3) Indian Institute of Science Education and Research, Bhopal 462 023, India}
\ec

\bc\begin{minipage}{0.9\textwidth}\begin{spacing}{1}{\small {\bf Abstract:}
We consider a model for gravity that is invariant under global scale transformations.
It includes one extra real scalar field coupled non-minimally to the gravity fields. In this model all the dimensionful parameters like the
gravitational constant and the cosmological constant etc. are generated by
a solution of the classical equations of motion. Hence
this solution provides a mechanism to break scale invariance.
 In this paper we demonstrate the 
stability of such a solution against small perturbations in a flat FRW background by making
a perturbative expansion around this solution and solving the resulting equations linear in the
perturbations.
This demonstrates the robustness of this symmetry breaking mechanism.
}\end{spacing}\end{minipage}\ec


\section{Introduction}
Scale invariance is an idea with quite a long history. 
The possibility of local scale invariance was first suggested by Weyl
\cite{Weyl:1929} in the twenties. 
This subsequently
lead to considerable research effort by several physicists
\cite{Deser,Dirac1973,Sen:1971,Utiyama:1973,Freund:1974,Hayashi:1976,Nishioka:1985,Padmanabhan85,ChengPRL,Hochberg91,Wheeler98,Pawlowski99,Nishino2009,Demir2004,Huang1989,Wei2006,JMS,AJS,Moon}.
Quantizing a scale invariant theory is considered problematic since scale invariance is anomalous in general. However, it has
also been argued that in some cases it might be possible to preserve scale invariance in the full quantum field
theory
\cite{Englert:1976,JMS,Shaposhnikov:2008a,JM09}. Despite the difficulties scale invariance remains an interesting idea since it
has the potential to resolve one of the
greatest puzzle of physics, namely, the cosmological constant problem\cite{JMS,Mannheim09,JM09,JM10_1,JM10_2}.

A scale invariant theory contains no dimensionful
parameter in the action. Hence in any realistic theory scale
invariance has to be broken in order to agree with observations.
There exist several mechanisms to break
scale invariance \cite{Fujii74,Cooper,Wood92,Feoli98,JM,JMS,Finelli}.
One such mechanism for breaking scale invariance is to assume
the existence of a classical background cosmological solution
\cite{Cooper,JMS,JM,Finelli}.
This may be
demonstrated by including just one scalar field besides gravity. 
Consider the following action,
\ba
S=\int d^4 x \sqrt{-g}\left[\frac{\beta \chi^2}{8} R+ \frac{1}{2}D^\mu \chi D_\mu \chi -\frac{1}{4}\lambda\chi^4\right],
\ea
where $\chi$ is a real scalar field. This action is invariant under a global scale transformation,
\ba
\chi \rar \chi \Lm, \quad x \rar x/\Lm.
\ea
Here we have chosen the conventions followed in Refs.
\cite{Donoghue1,Donoghue2} where
the flat space-time metric takes the form $(1,-1,-1,-1)$ and
the curvature tensor and its contractions are defined as,
\ba
 R^\mu_{\nu\alpha\beta}&=& -\partial_{\beta}\Gamma^\mu_{\nu\alpha} +\partial_{\alpha}\Gamma^\mu_{\nu\beta} +\Gamma^\mu_{\gamma\alpha}\Gamma^\gamma_{\nu\beta}
 -\Gamma^\mu_{\gamma\beta}\Gamma^\gamma_{\nu\alpha},\nn\\
R_{\nu\beta} &=& R^\mu_{\nu\beta\mu} \hspace{1mm} ,\hspace{4mm} R = R_{\nu\beta}g^{\nu\beta}.
\label{eq:R_notation}
\ea
Now, we seek a constant solution of the equation of motion for the scalar field $\chi$, so that the term $\left(\bt\chi^2/8\right)$ in the action
generates the effective gravitational constant. Hence we may drop the terms containing the derivatives of the scalar field in its equation of motion to obtain
the following relation,
\ba
{\beta R\over 4} &=& \lambda\chi^2\,.
\label{eq:auxi}
\ea
Here $R$ represents the classical scale covariant curvature scalar, as
defined in Eq. \ref{eq:R_notation}.
We assume the FRW metric with the spatial curvature parameter $k=0$ and
scale parameter $a(t)$ where $t$ is the cosmic time.
We obtain a solution
to the classical equations,
\ba
\chi = \chi_0 = {M_{\rm PL}\over \sqrt{2\pi\beta}}\,,
\label{eq:chi_Mpl}
\ea
where $M_{\rm PL}$ is the Planck mass.
The FRW scale parameter is given by
\ba
a(t) = a_0e^{H_0t},
\label{eq:at}
\ea
where $H_0$ is the Hubble parameter.
This sets the curvature scalar as $R=12H_0^2$.
Hence we obtain the solution corresponding to the de Sitter space-time. The 
background field $\chi_0$, the Hubble constant $H_0$ and the curvature 
scalar $R$ are constant for this solution. One may of course obtain
more realistic solutions by adding contributions from dark matter
candidates as well as standard model particles \cite{AJS}. One may
also allow the scalar field $\chi$ to be time dependent \cite{Finelli}.
In such cases $R$ will no longer be independent of time. This will complicate
our analysis but will not be conceptually different from the simple case
we consider. Hence in this paper we consider this simple case only.

The solution generates both the gravitational constant and an effective
cosmological constant. 
It is interesting that an effective cosmological constant is generated,
despite the fact that scale invariance forbids this term in the action. 
This might be indicative of the fundamental nature of this parameter
\cite{dadhich}. 
We have argued in earlier papers that this mechanism
for breaking scale invariance is different from spontaneous symmetry
breaking. In the latter case one is interested in the behavior
of the ground state of the Hamiltonian of matter fields 
under the symmetry transformation. Hence one
 makes a quantum expansion around the
minimum of the potential. In the present case we seek a time dependent
solution and the minimum of the potential is not directly relevant.
The time dependence of the solution comes from the scale parameter, $a(t)$.
In earlier papers we have called this phenomenon cosmological symmetry
breaking \cite{JM,JMS} in order to emphasize its difference from spontaneous 
symmetry breaking.
The important question now is whether such a solution is stable or not.
In this paper we investigate this question. We demonstrate the stability
of this solution under small perturbations and hence show that it can
consistently break scale invariance. 

Such de Sitter or approximately de Sitter solutions have also been
discussed in the case of general scalar-tensor theories. Several papers
have addressed the issue of stability of these solutions \cite{V_Faraoni1,V_Faraoni,QLN} 
as well as computed 
the power spectrum of perturbations \cite{Zhang_Li,J_Whang,CFTV,Bassett,Langlois,Durrer,Robert}. The power spectrum
is useful if one assumes that this solution is applicable to inflation. The
theory we consider is obtained as a special case of these
scalar-tensor theories by imposing the constraint that the action must
be invariant under scale transformations. The background de Sitter solution
plays a crucial role in our case since it provides a mechanism to break 
scale symmetry. Indeed it is difficult to find alternate methods to break
this symmetry. For example, this symmetry can be broken spontaneously only
if we arbitrarily set some terms to zero in the action 
\cite{Shaposhnikov:2008a}. Since
no symmetry prohibits such terms, these have to tuned to zero at each order
in perturbation theory. In contrast our mechanism does not require any
such constraint. Due to the special role played by the de Sitter solution
in our model, testing the stability of this solution is crucial in our
case. If the solution is found to be stable then it establishes the
cosmological symmetry breaking as a robust mechanism to break
scale and possibly other symmetries.   
   
We may also directly use the results of stability
analysis of the de Sitter solutions in general scalar-tensor theories
and apply these to our special case. 
A general criteria for testing stability is given, for example, in Ref. 
\cite{V_Faraoni1,V_Faraoni}.
We find that our results are consistent with this condition,
as discussed in Section 2.  

The background de Sitter solution in our scale invariant model may be
applicable to inflation or to dark energy. However in the present
paper we are not interested in cosmological applications of this model
and only in demonstrating that the solution is stable.

In the rest of this paper we shall use the conformal time $\eta$ instead of $t$. Hence the spatially flat FRW metric becomes,
\ba
g_{\m\n} = a^2(\et)(1,-1,-1,-1)\,.
\ea
In terms
of $\eta$ the Einstein's equations, at the leading order, may be written as,
\ba
\left(\frac{a'}{a}\right)^2&=&\frac{\lambda\chi_0^2a^2}{3\beta} \ ,\nn\\
\left(\frac{a''}{a}\right)&=&2\left(\frac{\lambda\chi_0^2a^2}{3\beta}\right)
\label{leading} \ .
\ea
Here the primes represent derivatives with respect to $\eta$.

\section{Perturbations}
In this section we study the perturbations to the leading order solution.
As already mentioned we shall restrict ourselves to small perturbations.
The perturbed metric may be expressed as,
\ba
g_{\mu\nu} = a^2\left( \begin{array}{cc}
1+2A & \partial_i B + M_i\\
\partial_i B + M_i& \left(-1 + 2\psi\right)\delta_{ij}-D_{ij}E +\partial_j V_i +\partial_i V_j + P_{ij} \end{array} \right) \ ,
\ea
where $a=a(\eta)$ is the scale factor of the universe in terms of the conformal
time, $D_{ij}=\left(\partial_i\partial_j-\frac{1}{3}\delta_{ij}\nabla^2\right)
$, $A, B, E$ and $\psi$ are the scalar perturbations, $V_i$ and $M_i$ are the
 vector perturbations and $P_{ij}$ stands for the pure tensor perturbations.
The pure vector and the tensor parts satisfy the following constraints \cite{Mukhanov},
\ba
\partial_iV_{i}=\partial_iM_{i}=0;\ \ \partial_i P_{ij}=0; \ \ P_{ii}=0 \label{Decomposition}.
\ea
 The inverse metric becomes,
\ba
g^{\mu\nu} = \frac{1}{a^2}\left( \begin{array}{cc}
1-2A &\partial_i B +M_i\\
\partial_i B +M_i & \left(-1 - 2\psi\right)\delta_{ij}+D_{ij}E -\partial_j V_i -\partial_i V_j - P_{ij} \end{array} \right),
\ea
where we have neglected all the terms beyond the first order in perturbations.
We also expand the field $\chi$ such that
\ba
\chi = \chi_0 + \hat{\chi},
\label{eq:chi_exp}
\ea
where $\chi_0$ represents the solution at leading order
and $\hat\chi$ a small perturbation.

We express the Einstein equation as,
\ba
G_{\mu\nu}&=&T_{\mu\nu} \ ,
\ea
where,
\ba
G_{\mu\nu}&=&R_{\mu\nu}-\frac{1}{2}g_{\mu\nu}R,\\
T_{\mu\nu}&=&-\frac{\chi^2_{;\lambda;\kappa}}{\chi^2}\left(g^\lambda_\mu g^\kappa_\nu-g_{\mu\nu} g^{\lambda\kappa}\right)
+\frac{4}{\beta\chi^2}\left(\frac{1}{2}g^{\rho\sigma}(\partial_\rho \chi)(\partial_\sigma \chi)
-\frac{1}{4}\lambda\chi^4\right)g_{\mu\nu}\nonumber\\&-&\frac{4}{\beta\chi^2}(\partial_\mu\chi)(\partial_\nu\chi) \ .
\ea
Using the decomposition theorem \cite{Norbert,Dodelson,Robert}, we can now
decompose  the
Einstein's equation into scalar, vector and tensor modes and treat their perturbations separately as
these modes evolve independently \cite{Lifshitz_1,Lifshitz_2,Bertschinger}.
 The tensor modes are gauge invariant \cite{Bardeen} and so gauge fixing is not required. However 
both in the case of the scalar and the vector modes we
need to fix a gauge to get a  physical solution.

\subsection{Scalar modes:}
For the scalar modes we choose the longitudinal (conformal Newtonian) gauge 
 \cite{Ma_Bertschinger,Riotto:2002,Weinberg_2}, $ B=E=0$.
At the first order in the perturbations, the components of the Einstein tensor are given by,
\begin{eqnarray}
\delta G_{00} &=& -2 \nabla^2\psi+6\frac{a'}{a}\psi' \ ,\\
\delta G_{0i}&=&-2\frac{a'}{a} \partial_i A - 2\partial_i \psi' \ ,\\
\delta G_{ij}&=&\left(-\nabla^2\left(A-\psi\right)-2\frac{a'}{a}A'-4\frac{a'}{a}\psi'-2\psi'' -\frac{2\lambda\chi_0^2a^2}{\beta}\left(A+\psi\right)\right)\delta_{ij}\nn\\
&&+\partial_i\partial_j(A-\psi) \ .
\end{eqnarray}
The corresponding components of the energy momentum tensor are
\begin{eqnarray}
\delta T_{00}&=&-2\nabla^2\left(\frac{\hat{\chi}}{\chi_0}\right)+6\left(\frac{a'}{a}\right)\left(\frac{\hat\chi'}{\chi_0}\right)-\frac{2\lambda\chi_0^2 a^2}{\beta}\left(A+\left(\frac{\hat{\chi}}{\chi_0}\right)\right) \ , \\
\delta T_{0i}&=&2\left(\frac{a'}{a}\right)\partial_i\left(\frac{\hat{\chi}}{\chi_0}\right)-2\partial_i\left(\frac{\hat{\chi'}}{\chi_0}\right) \ , \\
\delta T_{ij}&=&\left[2\nabla^2\left(\frac{\hat{\chi}}{\chi_0}\right)-2\left(\frac{a'}{a}\right)\left(\frac{\hat{\chi}'}{\chi_0}\right)-2\left(\frac{\hat{\chi}''}{\chi_0}\right)+\frac{2\lambda\chi_0^2a^2}{\beta}\left(\frac{\hat{\chi}}{\chi_0}-
\psi\right)\right]\delta_{ij}\nn\\
&&-2\partial_i\partial_j\left(\frac{\hat{\chi}}{\chi_0}\right).
\end{eqnarray}
Comparing all the components of the Einstein tensor and the energy momentum tensor we get,
\ba
- \nabla^2\psi+3\frac{a'}{a}\psi'&=&-\nabla^2\left(\frac{\hat{\chi}}{\chi_0}\right)+3\left(\frac{a'}{a}\right)\left(\frac{\hat\chi'}{\chi_0}\right)\nonumber\\
&&-\frac{\lambda\chi_0^2 a^2}{\beta}\left(A+\left(\frac{\hat{\chi}}{\chi_0}\right)\right),\label{xi_eqn_1}\\
-\frac{a'}{a} \partial_i A - \partial_i \psi'&=&\left(\frac{a'}{a}\right)\partial_i\left(\frac{\hat{\chi}}{\chi_0}\right)-\partial_i\left(\frac{\hat{\chi'}}{\chi_0}\right),\label{xi_eqn_2}\\
-\nabla^2\left(A-\psi\right)-3 \frac{a'}{a}A'-6 \frac{a'}{a}\psi'-3 \psi''&=&2\nabla^2\left(\frac{\hat{\chi}}{\chi_0}\right)-3\Big[ \left(\frac{a'}{a}\right)\left(\frac{\hat{\chi}'}{\chi_0}\right)+
\left(\frac{\hat{\chi}''}{\chi_0}\right)\nonumber\\
&&-\frac{\lambda\chi_0^2a^2}{\beta}\left(\frac{\hat{\chi}}{\chi_0}-\psi\right)+\frac{\lambda\chi_0^2a^2}{\beta}\left(A+\psi\right)\Big] \ ,\label{xi_eqn_3}\\
\partial_i \partial_j \left(A-\psi\right)&=&-2\partial_i \partial_j\left(\frac{\hat\chi}{\chi_0}\right)\ .\label{xi_eqn_4}
\ea
Since we seek solutions whose spatial dependence is equal to $e^{i\vec k\cdot \vec x}$,
from Eq. \ref{xi_eqn_4} we get,
\ba
A-\psi&=& -2\left(\frac{\hat{\chi}}{\chi_0}\right).\label{xi_eqn_5}
\ea
Using  Eq. \ref{leading} we find,
\ba
{a'\over a} = - {1\over \eta + C} \label{leading1} \ ,
\ea
where $C$ is a constant. Using  Eqs. \ref{xi_eqn_1} and  \ref{xi_eqn_5}
we obtain the following
solutions for $\xi$, defined as $\xi=\psi-\frac{\hat{\chi}}{\chi_0}$,
\begin{eqnarray}
 \xi&=&D(\eta+C)e^{\frac{k^2}{3}(\frac{\eta^2}{2}+C\eta)}\label{xi_soln_1}\label{xi_sol_1} \\ \text{or}\ \xi&=&0\label{xi_sol_2} \ ,
\end{eqnarray}
where $D$ is an arbitrary constant. Since the non-zero solution does not satisfy Eqs. \ref{xi_eqn_2} and  \ref{xi_eqn_3}, $\xi=0$ is the only
simultaneous solution to Eqs. \ref{xi_eqn_1}-\ref{xi_eqn_4}. This implies
\ba
A=-\psi=-\frac{\hat{\chi}}{\chi_0}\label{relation_A_psi} \ .
\ea
The background scale factor  $a(\eta)$  may be obtained by solving Eq. \ref{leading},
\ba
a(\eta) = -{1\over H_0(\eta + C)} \ .
\ea
Without loss of generality we may set $a(\eta=0) = 1$. This gives
\ba
C = -{1\over H_0}= -\eta_0.
\ea
As $a\rightarrow \infty$ for $\et \rar \et_0 $, the maximum value of $\eta$ is $\eta_0 = 1/H_0$. 

\begin{figure}[t!]
\begin{center}
\hspace{-3.5cm}\includegraphics [angle=0,width=1\linewidth] {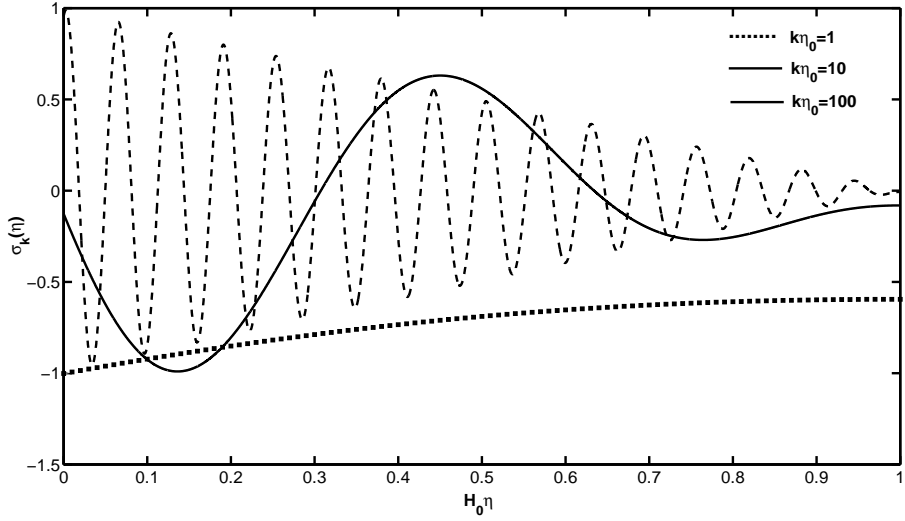}
\caption{The variation of $\sigma_{k}(\eta)$ for three different values of
 $k$.}
\label{perturbations}
\end{center}
\end{figure}

The equation of motion of $\hat{\chi}$ is given by,
\ba
-(1+2\beta)\nabla^2\left(\frac{\hat{\chi}}{\chi_0}\right)-
 \frac{\beta}{2}\nabla^2 A + \left(1+3 \beta \right)\left(\frac{\hat{\chi}''}{\chi_0}\right)
 + \frac{3 \beta}{2}\left(A''\right)&&\nonumber \\
+ \left(2+9\beta\right)\left(\frac{a'}{a}\right)\left(\frac{\hat{\chi}'}{\chi_0}\right)
+6\beta\left(\frac{a'}{a}\right)A'+2\lambda\chi_0^2 a^2\left(A+\left(\frac{\hat{\chi}}{\chi_0}\right)\right)&=&0 \ .
\ea
By using Eq. \ref{relation_A_psi} this equation reduces to,
\ba
\hspace{-3cm} \nabla^2 A -A'' - 2\left(\frac{a'}{a}\right)A'=0 \label{Diff_A} \ .
\ea
Let $A(\vec x,\eta)=\sigma_{k}(\eta) e^{i\vec k\cdot\vec x}$. We find
\ba
\sigma_{k}''(\eta)+ \frac{2 \sigma_{k}'(\eta)}{\eta_0 - \eta}= -k^2\sigma_{k}(\eta)   \  .
\label{eq:stability}
\ea
This equation is consistent with the condition given in Ref. \cite{V_Faraoni,V_Faraoni1}, expressed in terms of conformal time.
It corresponds to the limiting case of Eq. 30 of Ref. \cite{V_Faraoni}, 
where we use the equality sign. 
This is just the equation of a damped harmonic oscillator with the damping
force increasing with time. We point out that the maximum value of the 
conformal time $\eta$ is $\eta_0$. As $\eta\rightarrow \eta_0$, the background
scale factor approaches infinity. Hence the damping force is always positive
and we expect no unstable modes. 
The solution, $\sigma_k(\eta)$, is given by
\ba
 \sigma_k(\eta) &= & C_1\left[\sin k(\eta-\eta_0)-k(\eta-\eta_0)\cos k(\eta-\eta_0)\right]\nonumber\\
&-&C_2\left[\cos k(\eta-\eta_0)+k(\eta-\eta_0)\sin k(\eta-\eta_0)\right]\ .\label{eq:Axyzn}
\ea
where $k=|\vec k|$ and $C_1$ and $C_2$ are constants. 
The solution for some representative values of $k$ 
is shown  in the Fig. \ref{perturbations}. As expected we do not find
any unstable modes. We do, however, find a mode of zero frequency which  
arises if $k=|\vec k|=0$. In this case we find a solution,
$\sigma_0 = {\rm constant}$,
which is independent of
$\eta$. 

Our results show that the classical de Sitter solution, which breaks scale
invariance, is stable under small perturbations. 
Hence we have demonstrated that the cosmological symmetry breaking mechanism
consistently breaks
scale invariance. This is the main new result of our paper.
\subsection{Vector modes:}
We next consider the vector perturbations. 
For simplicity we rewrite the Einstein equation in the following form,
\ba
R_{\mu\nu}= T_{\mu\nu} - \frac{1}{2}g_{\mu\nu}T^{\alpha}_{\alpha} .\label{relation}
\ea
The $j-i$ components of Eq. \ref{relation} give,
\ba
\partial_j\left[V_i''+2\frac{a'}{a}V_i'-M_i'-2\frac{a'}{a}M_i\right]=0 \label{relation_vector1} 
\ea
and  $0-i$ components may be written as, 
\ba
\nabla^2(M_i-V_i')=0 \ .
\ea
Here we have used the Eq. \ref{Decomposition}.
We make a gauge choice as $M_i =0 $ which implies $\nabla^2 V_i = 0$. From Eq. \ref{relation_vector1} we get
\ba
\partial_j\left[V_i''+2\frac{a'}{a}V_i'\right]=0. \label{vector1}
\ea
As we seek solutions whose spatial dependence is given by $\exp(i\vec k\cdot \vec x)$ this implies,
\ba
  V_i''-\frac{2}{\eta + C}V_i' &=& 0,
\ea
where we have used Eq. \ref{leading1}.
Hence the solution of vector mode perturbation is
\ba
 V_i  &=& C^V_i \frac{(\eta -\et_0)^3}{3} e^{i\vec k\cdot \vec x} \ ,
\ea
where $C^V_i$ is a constant vector. This implies that as $\et \rar \et_0$, the vector perturbation dies off as $(1-\et/\et_0)^3$.

\subsection{Tensor modes:}
We finally consider the tensor modes.
As already mentioned we do not need to make any gauge choice to solve for the tensor mode.
Using Eq. \ref{Decomposition} the equation for the tensor modes can be written as,
\ba
\nabla^2 P_{ij}-2\frac{a'}{a}P_{ij}'-P_{ij}''=0 \ .
\ea
Comparing this with Eq. \ref{Diff_A} we see that $P_{ij}$ satisfies the same equation as $A$. Hence, the solution for the tensor modes
is same as that of the scalar perturbations.

\section{Conclusion}
We have considered a scale invariant model of gravity including an extra real scalar field. The scaling symmetry is broken by a mechanism which has some 
similarity to spontaneous symmetry breaking. In this case a time 
dependent solution of the classical equations of motion breaks the 
symmetry and 
generates all the dimensionful parameters of the theory like the gravitational constant and an effective cosmological constant. In this paper
we have analyzed the stability of such a solution against small perturbations.
We have shown that the symmetry breaking solution is stable against small
perturbations, irrespective of their 
scalar, vector or tensor nature. 
However the exact behavior depends on the nature of the perturbation. 
The scalar and tensor perturbations show damped oscillations as a function 
of conformal time $\eta$. 
The vector perturbation, however, decay monotonically with conformal time.
Our results establish the robustness of this mechanism to break scale 
invariance.


\begin{spacing}{1}
\begin{small}

\end{small}
\end{spacing}
\end{document}